\documentclass[aps,prl,reprint,superscriptaddress,amsmath,footinbib]{revtex4-1}
\usepackage[colorlinks=true,citecolor=blue,urlcolor=blue,linkcolor=blue]{hyperref}
\usepackage{graphicx}
\usepackage{braket}
\usepackage[normalem]{ulem}
\graphicspath{{../../Figures/}}
\DeclareGraphicsExtensions{.pdf,.png}

\begin{document}
\title{Interacting topological frequency converter}

\author{Simon K\"{o}rber}
\email[Email:\,]{skoerber@physik.uni-wuerzburg.de}
\affiliation{Institute of Theoretical Physics and Astrophysics, University of W\"{u}rzburg, Am Hubland, 97074 W\"{u}rzburg, Germany}

\author{Lorenzo Privitera}
\affiliation{Institute of Theoretical Physics and Astrophysics, University of W\"{u}rzburg, Am Hubland, 97074 W\"{u}rzburg, Germany}

\author{Jan Carl Budich}
\affiliation{Institute of Theoretical Physics, Technische Universit\"{a}t Dresden, 01062 Dresden, Germany}
\affiliation{W\"{u}rzburg-Dresden Cluster of Excellence ct.qmat, Germany}

\author{Bj\"orn Trauzettel}
\affiliation{Institute of Theoretical Physics and Astrophysics, University of W\"{u}rzburg, Am Hubland, 97074 W\"{u}rzburg, Germany}
\affiliation{W\"{u}rzburg-Dresden Cluster of Excellence ct.qmat, Germany}

%\date{\today}

\begin{abstract}
We show that an interacting two-spin model subjected to two circularly polarized drives enables a feasible realization of a correlated topological phase in synthetic dimensions. The topological observable is given by a quantized frequency conversion between the dynamical drives, which is why we coin it the \textit{interacting topological frequency converter} (ITFC). The ITFC is characterized by the interplay of interaction and synthetic dimension. This gives rise to striking topological phenomena that have no counterpart in the noninteracting regime. By calculating the topological phase diagrams as a function of interaction strength, we predict an enhancement of frequency conversion as a direct manifestation of the correlated topological response of the ITFC.
%\\[-13pt]\begin{center}{\it version~\input{|"git describe --tags"}}\\[-13pt]\end{center}
%\\[-13pt]\begin{center}{\it version~\input{version}}\\[-13pt]\end{center}
\end{abstract}


%\keywords{}

\maketitle

%he quest for the experimental realization and control of topological states of matter has greatly stimulated the study of non-equilibrium quantum systems~\cite{Hosur2011,McIver2012,Dora2012,Wang2013,Seetharam_PRX15,Budich_PRA15,Iadecola_PRB15,McGinley_PRB19,Dehghani_PRB16,Eckardt2017,Weinberg_PR17,Esin_PRB18,Reimann_Nature18,McIver2019,Sato_PRB19,Barbarino_arxiv19}.
{\it Introduction.}---Recent advances in the realization and manipulation of quantum systems far from thermal equilibrium have triggered the quest for identifying and observing topological phenomena in such settings~\cite{Goldman_NPhys16,Eckardt2017,Cooper_RMP19,Rudner_arxiv19}. In particular, it has become clear that periodic driving in time can induce intriguing topological features~\cite{Yao2007,Oka2009,Inoue2010,Lindner2011,Kitagawa2011,Gu2011,Jiang2011,Liu2012,Morell2012,Cayssol2013,Wang2013,Liu_PRL13,Iadecola2013,Thakurathi2013,Jotzu_Nature14,Privitera2016,Yan_PRL16,Roy2017,Roy2017b,Flaschner_Nphys18,Fleckenstein2019,RodriguezMena2019,Mera_PRL19}, numerous of which have been shown to be unique to Floquet systems~\cite{Kitagawa2010,Rudner2013,Asboth2014,Potter2016,Titum2016,Po2017,Budich_PRL17,Nathan_PRL17,Kolodrubetz2018,Sun_PRL18,Higashikawa_PRL19}. Many of these novel nonequilibrium phases rely on the fact that a time-periodic driving may formally be viewed as a \textit{dimensional extension} of the system~\cite{Grifoni1998,GomezLeon2013}. The archetypal example in this context is provided by the Thouless pump in one spatial dimension, which maps onto an integer quantum Hall scenario upon interpreting time as an extra dimension~\cite{Thouless1983,Lohse2016,Nakajima2016}. This line of reasoning can be generalized to multifrequency driving~\cite{Casati_PRL89,Grifoni1998,Ozawa_NPR19}. There, the harmonics of different drives are viewed as lattice sites along different \textit{synthetic} spatial dimensions, which allows for the realization of a wide range of topological effects~\cite{Martin_PRX17,Peng2018,Lohse2018,Nathan2019,Crowley2019,Peng_PRB19,Crowley_arxiv2019}. Along these lines, it has been shown in Refs.~\cite{Martin_PRX17, Crowley2019} that driving a qubit with two fields of incommensurate temporal periodicity generates a dynamical analog of a Chern insulator~\cite{Qi2006,Haldane1988}, where the Hall response translates into a quantized \textit{frequency conversion} between the external fields. 

\begin{figure}[t!]
	\includegraphics[width=\columnwidth]{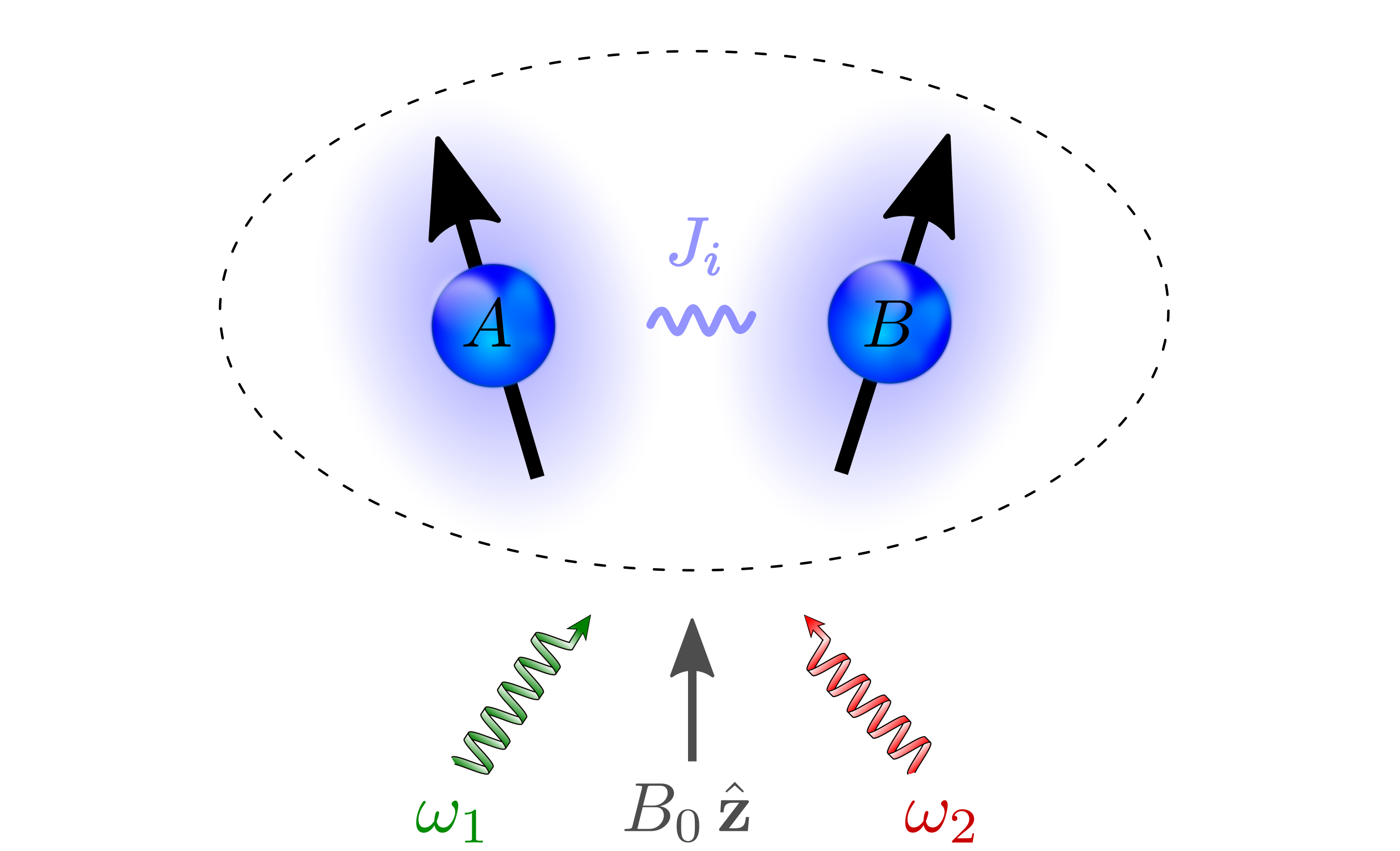}
	\caption{Interacting topological frequency converter (ITFC). The spins $A/B$ coupled by spin-spin interaction are subjected to a static magnetic field with amplitude $B_0$ and two circularly polarized drives with frequencies $\omega_{1}$ and $\omega_{2}$. The interaction is controlled by the coupling parameters $J_i$ $\{i=x,y,z\}$.}
	\label{fig:Setup}
\end{figure}

In this work, we extend the notion of topologically quantized frequency conversion to interacting spin systems. We focus on a minimal model of two interacting spins exposed to two circularly polarized incommensurate drives (see Fig.~\ref{fig:Setup} for an illustration). Despite its simplicity, this setup already offers a striking example of how the interplay of interaction and the aforementioned dynamical dimensional extension can have a profound impact on the resulting topological properties. Most prominently, while in a noninteracting system of two identical spins the topological charge determining the frequency conversion is constrained to be even, in the interacting case also odd integers are allowed. This feature may, in turn, result in an \textit{enhancement by interactions} of the topological response. For these reasons, we call our setup an \textit{interacting topological frequency converter} (ITFC). We provide a simple interpretation of our results in terms of two-body spin configurations and corroborate the observed topological phase diagram with an explicit calculation of the system's dynamics and the related frequency conversion. 

\begin{figure*}
	\centering
	\includegraphics[width=\textwidth]{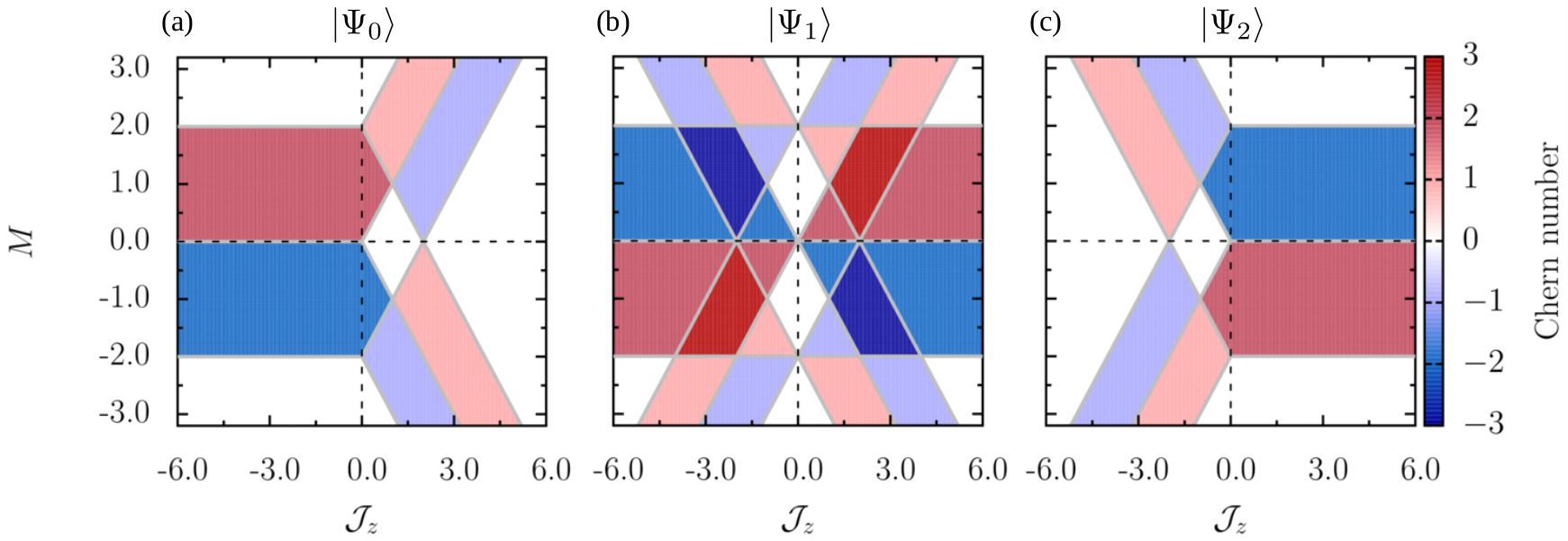}
	\caption{Topological phase diagrams for the eigenstates $\ket{\Psi_n(\vec{\varphi})}$ $\{n=0,1,2\}$ as a function of mass parameter $M$ and effective interaction strength $\mathcal{J}_z$ ($\mathcal{J}_{x-y}=0$). Due to many-body effects, interactions can drive the system into correlated topological phases with odd Chern numbers $C_n=\pm1,\,\pm3$. Furthermore, a finite interaction strength $\mathcal{J}_z\neq0$ can induce a nontrivial topology for states that were trivial in the noninteracting regime.}
	\label{fig:PhaseDiagram_Jz}
\end{figure*}

{\it Model.}---We analyze the ITFC schematically as shown in Fig.~\ref{fig:Setup}, in which two spins coupled by spin-spin interaction are exposed to a dynamical magnetic field $\mathbf{B}(\vec{\varphi}_t)$. The corresponding Hamiltonian is given by
\begin{equation}
\hat{H}(\vec{\varphi}_t)=\frac{g\,\mu_B}{2}\,\mathbf{B}(\vec{\varphi}_t)\,(\boldsymbol{\sigma}_A+\boldsymbol{\sigma}_B)+\sum_{i=x,y,z}J_i\,\sigma_{Ai}\,\sigma_{Bi},
\label{eq:FullHamiltonian}
\end{equation}
where we have introduced the vector of Pauli matrices $\boldsymbol{\sigma}_\alpha=(\sigma_{\alpha x},\sigma_{\alpha y},\sigma_{\alpha z})$ acting on the individual spins $\alpha=A/B$, and assumed anisotropic Heisenberg interaction with coupling parameters $J_i$ $\{i=x,y,z\}$. We focus on the case of spin 1/2, where $g=2$ is the bare $g$ factor of the electron and $\mu_B$ is the Bohr magneton. An experimental realization of such an interacting Hamiltonian could be implemented in gated double quantum dots~\cite{Petta_Science15,Bertrand2015}: in this case the bare $g$ factor is replaced by the effective factor $g^\ast$ (for GaAs $g^\ast=-0.44$). Alternatively, superconducting quantum circuits~\cite{Roushan2014} allow for the realization of this type of interaction, with a substantial degree of tunability of the anisotropy. In addition to a static magnetic field with amplitude $B_{0}$ in the $z$ direction, the external field  
\begin{equation}
\mathbf{B}(\vec{\varphi}_t)=\begin{pmatrix}
B_1\,\sin(\varphi_{1t})\\
B_2\,\sin(\varphi_{2t})\\
B_0-B_1\,\cos(\varphi_{1t})-B_2\,\cos(\varphi_{2t})
\end{pmatrix}
\label{eq:FieldVector}
\end{equation}
is composed of two circularly polarized drives with time-dependent phases $\vec{\varphi}_t=(\varphi_{1t},\varphi_{2t})=\vec{\omega}\,t+\vec{\phi}$ and amplitudes $B_{1/2}>0$, where the frequencies and offset phases are parametrized by $\vec{\omega}=(\omega_1,\omega_2)$ and $\vec{\phi}=(\phi_1,\phi_2)$. In the following, we set $B_{1/2}=B_c$ and $\hbar=1$ for simplicity.

The interaction favors ferromagnetic ($J_i<0$) or antiferromagnetic ($J_i>0$) alignment of the spins along the respective quantization axis in the ground state.  The interplay of interaction and magnetic field $\mathbf{B}(\vec{\varphi}_t)$  can be conveniently investigated by introducing the total spin $\hat{\mathbf{S}}=\frac{1}{2}\,(\boldsymbol{\sigma}_A+\boldsymbol{\sigma}_B)$ and the associated eigenstates $\ket{\psi_{s,m_z}}$ determined by the quantum numbers $(s,m_z)$: 
\begin{equation*}
\hat{\mathbf{S}}^{2}\ket{\psi_{s,m_z}}=s\,(s+1)\ket{\psi_{s,m_z}}, \quad \hat{S}_z\ket{\psi_{s,m_z}}=m_z\ket{\psi_{s,m_z}},
\end{equation*}
where $\{s=0,1\}$ and $\{m_z=-s,\dots ,s\}$. The Hamiltonian of Eq.~\eqref{eq:FullHamiltonian} commutes with the total spin $[\hat{H},\hat{\mathbf{S}}^{2}]=0$, so that the total spin quantum number $s$ is  conserved. Thus, the singlet state $\ket{\psi_{0,0}}$ is an eigenstate of the interacting system with trivial dynamics. For this reason, we restrict ourselves to study the Hilbert subspace with $s=1$. In this case, apart from a global constant, the Hamiltonian~\eqref{eq:FullHamiltonian} can be written as 
\begin{equation}
\hat{H}_T=\lambda\begin{pmatrix}
2\,\frac{B_z}{B_c}+\mathcal{J}_z && \sqrt{2}\,\frac{ B_-}{B_c} && \mathcal{J}_{x-y} \\
\sqrt{2}\,\frac{B_+}{B_c} && -\mathcal{J}_z && \sqrt{2}\,\frac{B_-}{B_c}\\
\mathcal{J}_{x-y} && \sqrt{2}\,\frac{B_+}{B_c} && -2\,\frac{B_z}{B_c}+\mathcal{J}_z
\end{pmatrix} \,,
\label{eq:TripletHamiltonian}
\end{equation}
in the basis of triplet states $\{\ket{\psi_{1,1}},\ket{\psi_{1,0}},\ket{\psi_{1,-1}}\}$. We have introduced the transverse components $B_\pm=B_x\pm i\,B_y$, the energy scale $\lambda=\frac{g\,\mu_B\,B_c}{2}$, and the effective interaction strengths $\mathcal{J}_{x\pm y}=\frac{J_x\pm J_y}{\lambda}$ and $\mathcal{J}_z=\frac{J_z}{\lambda}-\frac{\mathcal{J}_{x+y}}{2}$. The interaction enters $\hat{H}_T$ with the  parameters  $\mathcal{J}_z$ and $\mathcal{J}_{x-y}$, while the third parameter $\mathcal{J}_{x+y}$ only affects the energy $E_{0,0}=-\lambda\,(2\,\mathcal{J}_{x+y}+\mathcal{J}_z)$ of the decoupled singlet state $\ket{\psi_{0,0}}$. With isotropic exchange interaction, the interacting part of Eq.~\eqref{eq:FullHamiltonian} commutes with $\hat{S}_z$, effectively resulting in interaction strengths $\mathcal{J}_z=\mathcal{J}_{x-y}=0$. Below, we  focus on the anisotropic case, setting $\mathcal{J}_{x-y}=0$ for simplicity. Additional results for nonvanishing $\mathcal{J}_{x-y}$ are presented in the Appendix, showing that our main predictions are not qualitatively affected. 

\begin{figure*}[t!]
	\centering
	\includegraphics[width=\textwidth]{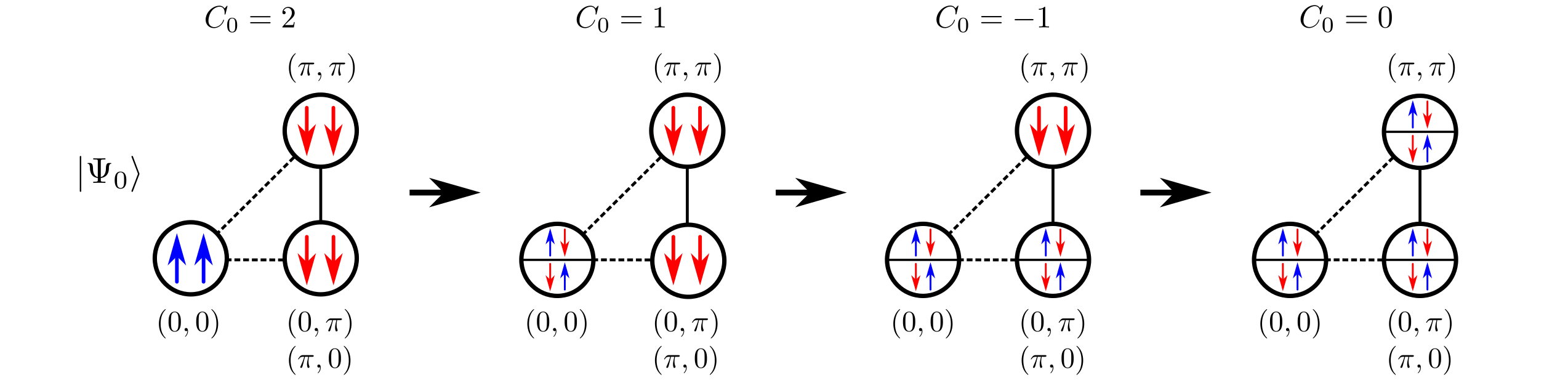}
	\caption{Spin configurations of the ground state $\ket{\Psi_0(\vec{\varphi})}$ at HSPs for interaction strength $\mathcal{J}_z\geq0$ and $M=1.2$ (see also Fig.~\ref{fig:PhaseDiagram_Jz}). Topological phase transitions caused by $\mathcal{J}_z$ lead to inversions between separable and maximally entangled states, which is why it is no longer possible to specify a winding number for the individual constituents. To describe the topology, we consider each separable state that contributes to the linear combination in $\ket{\psi_{1,0}}$ individually according to its topological features. Depending on the number $\{k=0,...,3\}$ of HSPs~\cite{Note1,Note2} with a maximally entangled state $\ket{\psi_{1,0}}$ as ground state, this leads to $2^k$ combinations for the spin configurations formed by the equally weighted quantum states $\ket{\uparrow\downarrow}$ and $\ket{\downarrow\uparrow}$ at different HSPs.}
	\label{fig:Winding}
\end{figure*} 

{\it Topological phase diagrams.}---For each qubit, the noninteracting part of Eq.~\eqref{eq:FullHamiltonian} is equivalent to a Chern insulator~\cite{bernevig2006quantum,Qi2006,Haldane1988} with mass parameter $M=B_0/B_c$. The time-dependent phases $\varphi_{1t}$ and $\varphi_{2t}$ play the role of Bloch quasimomenta: as they vary between $0$ and $2\pi$, they define a two-dimensional torus, analogous to a 2D Brillouin zone (BZ). Hence, the dynamics of the magnetic field $\mathbf{B}(\vec{\varphi}_t)$ of Eq.~\eqref{eq:FieldVector} can induce nontrivial topology in the single-spin subspaces for quasiadiabatic driving~\cite{Martin_PRX17}. Each of the two eigenstates of the single-spin Hamiltonian can be characterized by a Chern number: $\nu=\pm 1$ (nontrivial) for $|M|\leq 2$ or $\nu=0$ (trivial) for $|M|\geq 2$~\cite{Berry1984,Hasan2010}. In the interacting case, we can diagonalize the projected Hamiltonian in Eq.~\eqref{eq:TripletHamiltonian} and determine the Chern number~\cite{Xiao2010,thouless1982quantized,Bernevig2013} of the respective eigenstates $\ket{\Psi_n(\vec{\varphi})}$ $\{n=0,1,2\}$ according to
\begin{multline}
C_n=\frac{i}{2\,\pi}\int_{\text{BZ}}\mathrm{d}^2\vec{\varphi}\,[\braket{\partial_{\varphi_1}\Psi_n(\vec{\varphi})\,|\,\partial_{\varphi_2}\Psi_n(\vec{\varphi})}-\\
\braket{\partial_{\varphi_2}\Psi_n(\vec{\varphi})\,|\,\partial_{\varphi_1}\Psi_n(\vec{\varphi})}].
\label{eq:ChernNumber}
\end{multline}
The resulting topological phase diagrams as a function of mass parameter $M$ and effective interaction strength $\mathcal{J}_z$ ($\mathcal{J}_{x-y}=0$) are displayed in Fig.~\ref{fig:PhaseDiagram_Jz}. Interactions have two striking effects on the topology of the eigenstates. First, phases with odd Chern numbers $C_n=\pm1,\,\pm3$ emerge. This observation is a genuine interaction effect since for $\mathcal{J}_z=0$ the two qubits are independently exposed to the same magnetic field $\mathbf{B}(\vec{\varphi}_t)$, so that the global topological invariant can only change by an even number $\Delta C_n=\pm 2,\,\pm4$. Second, a finite interaction strength $\mathcal{J}_z\neq0$ can induce a nontrivial topology for states that were trivial in the noninteracting regime.

\begin{table}[t!]
	\centering
	\begin{tabular}{c|c}
		HSP	&  $\mathcal{J}_z$\\
		\hline
		$(0,0)$ & $\pm\,|2-M|$\\
		$(0,\pi),(\pi,0)$ & $\pm\,|M|$\\
		$(\pi,\pi)$ & $\pm\,|2+M|$
	\end{tabular}
	\caption{Interaction strength $\mathcal{J}_z$ (as a function of $M$) for which a Dirac gap closing happens at high-symmetry points (HSPs) of the BZ. Positive (negative) $\mathcal{J}_z$ correspond to inversions between the ferromagnetic ground (highest excited) state and the antiferromagnetic state $\ket{\psi_{1,0}}$. Since gap closings and reopenings at $(0,\pi)$ and $(\pi,0)$ occur simultaneously, the change in the global Chern number is twice as large as in the other HSPs.}
	\label{tab:BandInversion}
\end{table}

The topological phase transitions are caused by band inversions at high-symmetry points (HSPs) of the analog of the BZ, accompanied by Dirac gap closings for mass parameters $|M|=2$ or $M=0$, and effective interaction strengths $\mathcal{J}_z$ shown in Tab.~\ref{tab:BandInversion}. Notably, gap closings and reopenings at $(0,\pi)$ and $(\pi,0)$ occur simultaneously, so that the change in the global Chern number is twice as large as in $(0,0)$ or $(\pi,\pi)$. At HSPs, the Hamiltonian~\eqref{eq:FullHamiltonian} commutes  with $\hat{S}_z$, making $\hat{H}_T$ diagonal in the $\ket{\psi_{s,m_z}}$ basis: Gap closings caused by $\mathcal{J}_z$ ($M=\text{const.}$)  lead to inversions between the ferromagnetic triplet states $\ket{\psi_{1,1}}$, $\ket{\psi_{1,-1}}$ and the antiferromagnetic triplet state $\ket{\psi_{1,0}}$. The corresponding topological phases are bounded by straight lines in Fig.~\ref{fig:PhaseDiagram_Jz}. Topological phase transitions for mass parameters $|M|=2$ or $M=0$, however, only involve the ferromagnetic states $\ket{\psi_{1,1}}$ and $\ket{\psi_{1,-1}}$, which is why the polarization of both spins reverses upon band inversion at the respective HSP. The global topological invariant then changes by an even number $\Delta C_n=\pm 2,\,\pm4$. Conversely, a band inversion containing the antiferromagnetic state $\ket{\psi_{1,0}}$ is the reason for the generation of the odd topological phases with $C_n=\pm1,\,\pm3$.

{\it Interpretation of the results.}---From the previous considerations, it seems illuminating to investigate the relationship between the spin configuration of the eigenstates and the respective topological invariants. In our case, it is sufficient to focus on HSPs, which completely determine the topology of the system~\footnote{Since the Hamiltonian~\eqref{eq:FullHamiltonian} commutes with $\hat{S}_z$ at HSPs, there is a one-to-one mapping (see Fig.~\ref{fig:Winding}) from the spin configurations at HSPs to the topological invariant of Eq.~\eqref{eq:ChernNumber}. This applies whenever $\mathcal{J}_z\neq0$ and $\mathcal{J}_{x-y}=0$, while for nonvanishing $\mathcal{J}_{x-y}$ this argument breaks down. The topological invariant then has to be calculated by Eq.~\eqref{eq:ChernNumber}, but as discussed in the Appendix, this does not qualitatively change the phenomenological results of the work.}. In Fig.~\ref{fig:Winding}, the spin configurations at HSPs are schematically shown for an increasing interaction strength $\mathcal{J}_z\geq0$ and fixed mass parameter $M=1.2$ (compare also to Fig.~\ref{fig:PhaseDiagram_Jz}). For $\mathcal{J}_z=0$, the ground state $\ket{\Psi_0(\vec{\varphi})}$ is always a  separable state, and the global topological features can be interpreted by examining the qubits separately: each one has a well-defined single-particle Chern number corresponding to the winding number in the single-qubit Bloch sphere as $\vec{\varphi}$ varies. The global Chern number $C_n$ is then simply the sum of the two single-particle ones. Since the two qubits at $(0,0)$ are polarized in opposite directions with respect to the other HSPs, each spin winds once around its Bloch sphere resulting in the combined Chern number $C_0=2$.  

By increasing the interaction strength $\mathcal{J}_z>0$, a phase transition into the correlated topological phase $C_0=1$ is achieved. Within this phase, the separable ground state at $(0,0)$ is substituted by the maximally entangled ground state $\ket{\psi_{1,0}}$. Thus, since it is no longer possible to specify a winding number for the individual constituents, we exploit the following idea: We examine the topological features of each separable state that contributes to the linear combination in $\ket{\psi_{1,0}}$ individually. For instance, for the phase of $C_0=1$ the topology for the equally weighted quantum states $(\text{a})\,\ket{\uparrow\downarrow}^{(0,0)}$ and $(\text{b})\,\ket{\downarrow\uparrow}^{(0,0)}$ at $(0,0)$ is examined. Each separable state shows a single-spin winding number $\nu^{(\text{a})}_A=1$ or $\nu^{(\text{b})}_A=0$ for spin $A$. Moreover, spin $B$ is antiferromagnetically correlated to spin $A$, resulting in the winding number $\nu^{(\text{a})}_B=0$ or $\nu^{(\text{b})}_B=1$. In both cases $(\text{a})$ and $(\text{b})$, the single-particle Chern numbers add up to a global Chern number $C_0=1$. This picture provides an intuitive explanation for the odd topological phase: each band inversion between an antiferromagnetic (maximally entangled) state and a ferromagnetic (separable) state causes a change in the global Chern number by $\Delta C_n=\pm1$. 

When the interaction strength $\mathcal{J}_z>0$ is increased up to the phase $C_0=-1$, the separable ground states at $(0,\pi)$, $(\pi,0)$ are  replaced by the maximally entangled ground states $\ket{\psi_{1,0}}$. By applying the previous picture, we have to consider four~\footnote{Due to the symmetry of the Hamiltonian $\hat{H}(0,\pi)=\hat{H}(\pi,0)$, the quantum states at $(0,\pi)$, $(\pi,0)$ show the same spin configuration.} combinations for the spin configurations formed by the equally weighted quantum states $\ket{\uparrow\downarrow}$ and $\ket{\downarrow\uparrow}$ at different HSPs: $\ket{\uparrow\downarrow}^{(0,0)}\,\ket{\uparrow\downarrow}^{(0,\pi)}$, $\ket{\uparrow\downarrow}^{(0,0)}$ $\ket{\downarrow\uparrow}^{(0,\pi)}$, $\ket{\downarrow\uparrow}^{(0,0)}\,\ket{\uparrow\downarrow}^{(0,\pi)}$, and $\ket{\downarrow\uparrow}^{(0,0)}$ $\ket{\downarrow\uparrow}^{(0,\pi)}$. For each combination, the corresponding single-spin Chern numbers add up to the combined Chern number $C_0=-1$. This mechanism is quite generic and applies whenever $\mathcal{J}_z \neq 0$~\cite{Note1}: If $\{k=0,...,3\}$ is the number of HSPs~\cite{Note2} with a maximally entangled state $\ket{\psi_{1,0}}$, we have to examine $2^k$ equally weighted combinations of  spin configurations individually  (see Fig.~\ref{fig:Winding}). Topological transitions are then associated with a change in the configuration of the eigenstate under consideration. 

\begin{figure}[t!]
	\includegraphics[width=\columnwidth]{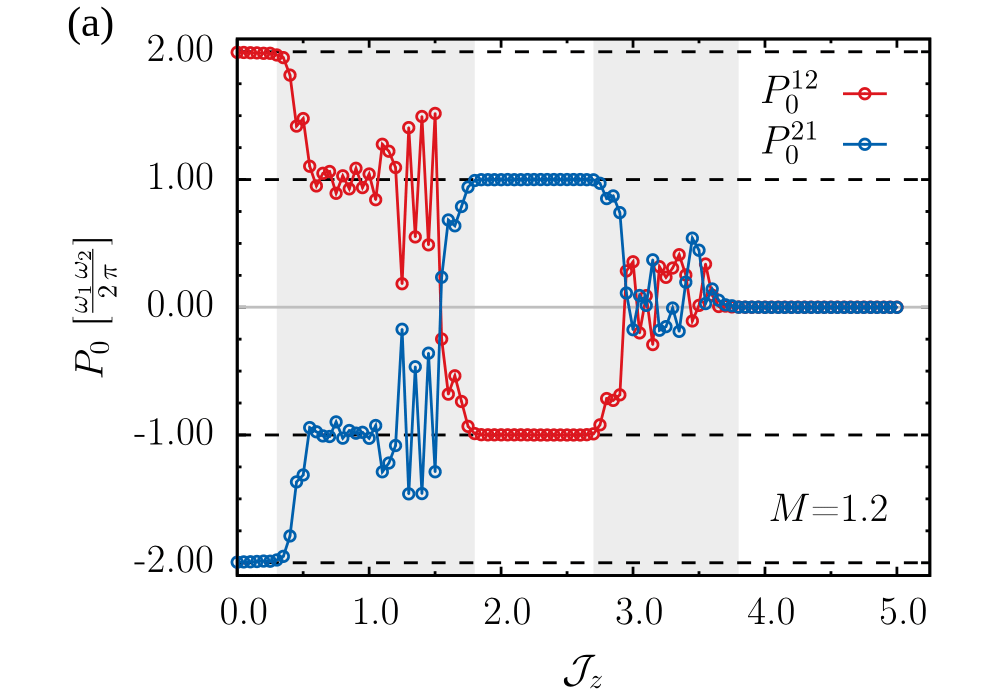}
	\includegraphics[width=\columnwidth]{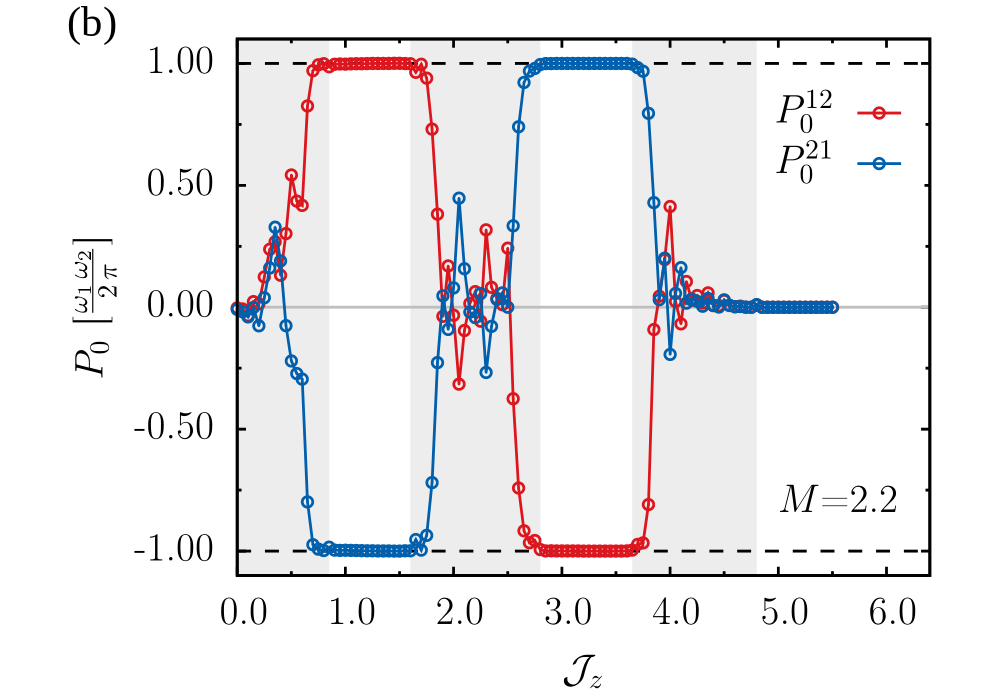}
	\caption{Extrapolated energy pumping rate $P_0$ as a function of $\mathcal{J}_z$ for parameters (a) $M=1.2$ and (b) $M=2.2$. The quantization occurs in an excellent agreement (white regimes) with the topological phase diagrams of Fig.~\ref{fig:PhaseDiagram_Jz}. This applies as long as the system shows quasiadiabaticity~\cite{Note3}, while otherwise the topological response breaks down (gray regimes).}
	\label{fig:EnergyPump_Jz}
\end{figure}

{\it Observable consequences.}---The dynamics of the system can also be described within a two-dimensional frequency lattice with built-in ''electric'' field~\cite{Ho1983,Verdeny2016,Martin_PRX17,Crowley2019}, where the Hall response in the nontrivial regime is given by a transverse current in this frequency domain. For a system initially prepared in an eigenstate of $\hat{H}_T$, the topological features of Fig.~\ref{fig:PhaseDiagram_Jz} then translate into a quantized frequency conversion between the circularly polarized drives~\cite{Martin_PRX17,Crowley2019}. As first realized in Ref.~\cite{Martin_PRX17}, energy is pumped between the two fields at a time-averaged rate
\begin{equation}
P_n^{12}=-P_n^{21}=\frac{C_n}{2\,\pi}\,\omega_1\,\omega_2
\label{eq:PumpingRate}
\end{equation} 
when the system is confined to an energy level of Chern number $C_n$. A necessary condition for the observation of a quantized rate, besides quasiadiabaticity~\footnote{Quasiadiabaticity is given for frequencies $\omega_1$ and $\omega_2$ that, on the one hand, are much smaller than the minimum band gap of the Hamiltonian~\eqref{eq:TripletHamiltonian} and, on the other hand, are large enough to produce a nonzero pumping rate~\eqref{eq:PumpingRate}. This ensures that over time interband excitations are suppressed.}, is that the two frequencies are incommensurate in such a way that the dynamics effectively samples the whole BZ. Notably, the enhancement of the frequency conversion for $C_n=\pm3$ is a direct consequence of the correlated topological response.

We investigate the topological energy pumping by numerically solving the Schr\"{o}dinger equation associated with $\hat{H}$~\footnote{We have multiplied~\eqref{eq:FullHamiltonian} by an overall factor $\eta=2.0$, in which formally increases the minimum band gap of the system and thus improves the ability of the quantum state to stay within the energy level during time evolution up to times $t=10^4\,\frac{1}{\lambda}$.}. We then compute  the expectation value of the ``current operator'' $\hat{\mathbf{J}}=\nabla_{\vec{\varphi}}\,\hat{H}$, which is associated with the total energy transfer rate as can be seen from the Heisenberg equation: $\frac{\mathrm{d}}{\mathrm{d}t}\braket{\hat{H}}=\braket{\partial_t\,\hat{H}}=\vec{\omega}\,\braket{\hat{\mathbf{J}}}$. If the initial state is the eigenstate $\ket{\Psi_n(\vec{\varphi}(t=0))}$, then the average pumping rate $P_n^{12}$ can be extrapolated through linear regression of the energy transfer $E_i(t)=\omega_i\,\int_{0}^{t}\mathrm{d}t'\,\braket{\hat{\mathbf{J}}_i(t')}$~\cite{Martin_PRX17}. We have chosen the frequencies $\omega_1=0.1\,\lambda$ and $\omega_2=\gamma\,\omega_1$, where $\gamma=\frac{1}{2}(1+\sqrt{5})$ is the golden ratio. The offset phases are $\phi_1=\pi/10$, $\phi_2=0$.  In Fig.~\ref{fig:EnergyPump_Jz}, the extrapolated pumping rate of the ground state is shown as a function of $\mathcal{J}_z$ for parameters $M=1.2$ and $M=2.2$. A quantized frequency conversion occurs in an excellent agreement (white regimes) with the topological phase diagrams of Fig.~\ref{fig:PhaseDiagram_Jz}. This applies as long as the ITFC shows quasiadiabaticity. Otherwise, the topological response is suppressed (gray regimes), and perfect quantization breaks down.

{\it Conclusion.}---We have demonstrated that combining two-body interaction with the notion of dynamically induced \textit{synthetic} dimensions gives rise to remarkable topological phenomena that go beyond a noninteracting implementation. In particular, we predict that an \textit{interacting topological frequency converter} (ITFC) realizes correlated topological phases with odd Chern numbers $C_n=\pm1,\,\pm3$. The topological response of the ITFC is given by a quantized \textit{frequency conversion} between the dynamical drives, and can be \textit{enhanced} for $C_n=\pm3$, as compared to its noninteracting counterpart. This amplification is found to be more pronounced as the number of interacting spins increases, as we explicitly confirmed in the Appendix for the example of three spins where Chern numbers up to $C_n=\pm5$ are observed. An experimental realization of the ITFC proposed in our present work might be implemented in a singlet-triplet qubit~\cite{Petta_Science15,Bertrand2015} exposed to circularly polarized driving fields or in superconducting quantum circuits~\cite{Roushan2014}.

\begin{acknowledgments}
This work was supported by the DFG (Grants No. SPP1666 and No. SFB1170 "ToCoTronics"), the W\"{u}rzburg-Dresden Cluster of Excellence ct.qmat (EXC2147, Project No. 39085490), and the Elitenetzwerk Bayern Graduate School on "Topological Insulators". We acknowledge Caterina De Franco for technical advices.
\end{acknowledgments}

\bibliographystyle{apsrev4-1}
%\bibliography{ThesisBiblio,FrequencyConverterNotes}

\onecolumngrid
\appendix
\section*{Appendix}

\subsection{Topological Phase Diagrams for nonvanishing $\boldsymbol{\mathcal{J}_{x-y}}$}
In the following, we investigate the influence of a nonvanishing interaction strength $\mathcal{J}_{x-y}$ on the topological phase diagrams of Fig.~\ref{fig:PhaseDiagram_Jz} of the main text. For a finite $\mathcal{J}_{x-y}\neq0$, the Hamiltonian~\eqref{eq:FullHamiltonian} does not commute with $\hat{S}_z$ at HSPs of the BZ, leading to a coupling of the ferromagnetic triplet states $\ket{\psi_{1,1}}$, $\ket{\psi_{1,-1}}$. A band inversion at HSPs then occurs between the antiferromagnetic triplet state $\ket{\psi_{1,0}}$ and a superposition of ferromagnetic triplet states $\ket{\psi_{1,1}}$ and $\ket{\psi_{1,-1}}$. Accordingly, the topology of the system can no longer be determined by considering the spin configurations at HSPs only, but has to be calculated by the full integral of Eq.~\eqref{eq:ChernNumber} of the main text. Thus, we have to diagonalize the projected Hamiltonian~\eqref{eq:TripletHamiltonian} for a finite $\mathcal{J}_{x-y}\neq0$, and calculate the Chern number of the respective eigenstates $\ket{\Psi_n(\vec{\varphi})}$ $\{n=0,1,2\}$ by Eq.~\eqref{eq:ChernNumber} of the main text. 

\begin{figure}
	\centering
	\includegraphics[width=\textwidth]{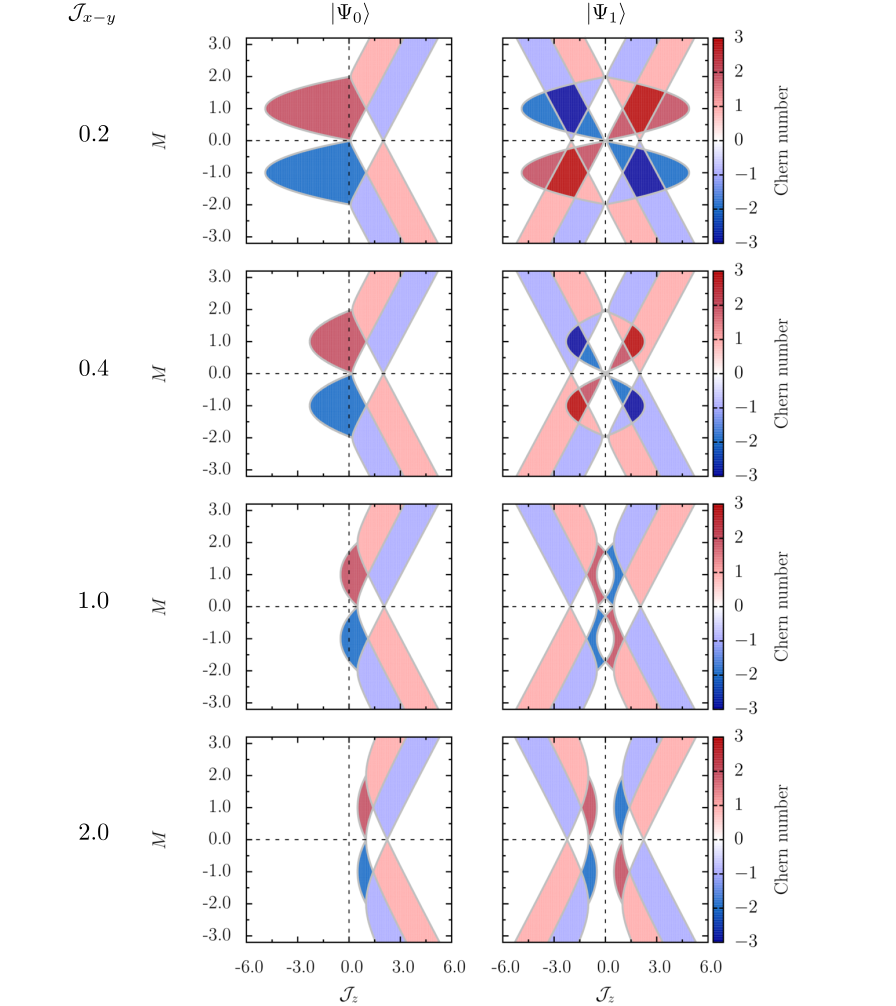}
	\caption{Topological phase diagrams as a function of $M$ and $\mathcal{J}_z$ for different interaction strengths $\mathcal{J}_{x-y}$. The topological phases are no longer bounded by straight lines as in Fig.~\ref{fig:PhaseDiagram_Jz} of the main text, which is why the topological phase transitions are quantitatively modified by a finite $\mathcal{J}_{x-y}\neq0$. The topological invariants of the phases, however, do not differ from those of the main text. In particular, the striking topological phenomena with odd Chern numbers $C_n=\pm1,\,\pm3$ are still present, and only influenced by a shift of the boundaries. The phenomenological results of the main text are therefore not qualitatively influenced by a finite interaction strength $\mathcal{J}_{x-y}$.}
	\label{fig:PhaseDiagram_Jz_Jxy}
\end{figure}

The resulting topological phase diagrams as a function of mass parameter $M$ and interaction strength $\mathcal{J}_z$ are displayed in Fig.~\ref{fig:PhaseDiagram_Jz_Jxy} for different interaction strengths $\mathcal{J}_{x-y}\neq0$. For a finite $\mathcal{J}_{x-y}$, the topological phases are no longer bounded by straight lines as in Fig.~\ref{fig:PhaseDiagram_Jz} of the main text. Especially, phase transitions that corresponded to the horizontal lines at $|M|=2$ and $M=0$ now show dispersive character. Although the critical values of $\mathcal{J}_z$ and $M$ are strongly modified by the coupling of the ferromagnetic triplet states, the corresponding topological invariants of the phases do not vary from those of the main text. In particular, the striking topological phenomena with odd Chern numbers $C_n=\pm1,\,\pm3$ are still present and are influenced at most only by a shift of the boundaries of the phase transitions. Accordingly, the phenomenological results of the main text are not qualitatively affected by a nonvanishing interaction strength $\mathcal{J}_{x-y}$.

%\newpage

\subsection{Topological Phase Diagrams for three interacting spins}
In the main text, we have demonstrated that the \textit{interacting topological frequency converter} (ITFC) consisting of two interacting spins (see Fig.~\ref{fig:Setup} of the main text for an illustration) realizes correlated topological phases in which especially the topological response can be \textit{enhanced by interaction} for global Chern numbers $C_n=\pm 3$. In the following, we indicate that such correlated topological phases are also to be expected with an increasing number of interacting spins, and that then the associated amplification of the topological frequency conversion is even more pronounced. As a prime example, we extend the Hamiltonian~\eqref{eq:FullHamiltonian} of the main text by an additional spin $C$ and, for simplicity, concentrate on a symmetrical arrangement of the spins given by
\begin{equation}
	\hat{H}_3(\vec{\varphi}_t)=\frac{g\,\mu_B}{2}\,\mathbf{B}(\vec{\varphi}_t)\,(\boldsymbol{\sigma}_A+\boldsymbol{\sigma}_B+\boldsymbol{\sigma}_C)+\sum_{i=x,y,z}J_i\,(\sigma_{Ai}\,\sigma_{Bi}+\,\sigma_{Bi}\,\sigma_{Ci}+\sigma_{Ai}\,\sigma_{Ci}).
\end{equation}
In the external field $\mathbf{B}(\vec{\varphi}_t)$ of Eq.~\eqref{eq:FieldVector} of the main text we again set the amplitudes to $B_{1/2}=B_c$. For the interacting three-spin model $\hat{H}_3$, the same calculations as in the main text for two spins can now be performed. In particular, the system again commutes with the total spin $[\hat{H}_3,\hat{\mathbf{S}}^{2}]=0$, and the problem can be decoupled into the Hilbert subspaces with $s=3/2$ and $s=1/2$. Apart from a global constant, the Hamiltonian can then be written in a diagonal form: $\hat{H}_3=\text{diag}[\hat{H}_Q,\hat{H}_{D1},\hat{H}_{D2}]$, where
\begin{equation}
	\hat{H}_Q=\lambda\begin{pmatrix}
		3\,(\frac{B_z}{B_c}+\mathcal{J}_z) && \sqrt{3}\,\frac{B_-}{B_c} && \sqrt{3}\,\mathcal{J}_{x-y} && 0 \\
		\sqrt{3}\,\frac{B_+}{B_c} && -\mathcal{J}_z+\frac{B_z}{B_c} && 2\,\frac{B_-}{B_c} && \sqrt{3}\,\mathcal{J}_{x-y} \\
		\sqrt{3}\,\mathcal{J}_{x-y} && 2\,\frac{B_+}{B_c} && -\mathcal{J}_z-\frac{B_z}{B_c} && \sqrt{3}\,\frac{B_-}{B_c} \\
		0 && \sqrt{3}\,\mathcal{J}_{x-y} && \sqrt{3}\,\frac{B_+}{B_c} &&     -3\,(\frac{B_z}{B_c}-\mathcal{J}_z)
	\end{pmatrix}
\end{equation}
is represented in the basis of quartet states $\{\ket{\psi_{\frac{3}{2},\frac{3}{2}}},\ket{\psi_{\frac{3}{2},\frac{1}{2}}},\ket{\psi_{\frac{3}{2},-\frac{1}{2}}},\ket{\psi_{\frac{3}{2},-\frac{3}{2}}}\}$. Again, we have introduced the transverse components $B_\pm=B_x\pm i\,B_y$, the energy scale $\lambda=\frac{g\,\mu_B\,B_c}{2}$, and the effective interaction strengths $\mathcal{J}_{x\pm y}=\frac{J_x\pm J_y}{\lambda}$ and $\mathcal{J}_z=\frac{J_z}{\lambda}-\frac{\mathcal{J}_{x+y}}{2}$ for ease of notation. The Hilbert subspace with $s=1/2$ consists of two degenerate doublet states $\{\ket{\psi_{\frac{1}{2},\frac{1}{2}}}^{(D1)},\ket{\psi_{\frac{1}{2},-\frac{1}{2}}}^{(D1)}\}$ and $\{\ket{\psi_{\frac{1}{2},\frac{1}{2}}}^{(D2)},\ket{\psi_{\frac{1}{2},-\frac{1}{2}}}^{(D2)}\}$ (we have chosen the states to form an orthonormal basis), each showing the structure (except for an overall energy shift caused by $\mathcal{J}_{x+y}$ and $\mathcal{J}_z$) of a noninteracting Chern insulator: 
\begin{equation}
	\hat{H}_{D1/D2}=\lambda\begin{pmatrix}
		\frac{B_z}{B_c} && \frac{B_-}{B_c} \\
		\frac{B_+}{B_c} && -\frac{B_z}{B_c}
	\end{pmatrix}-\lambda\,(3\,\mathcal{J}_{x+y}+\mathcal{J}_z).
\end{equation}
For this reason, we restrict ourselves to study the Hilbert subspace with $s=3/2$, for which we diagonalize the projected Hamiltonian $\hat{H}_Q$ for a finite interaction strength $\mathcal{J}_z\neq0$ ($\mathcal{J}_{x-y}=0$) and determine the Chern number of the respective eigenstates $\ket{\Psi_n(\vec{\varphi})}$ $\{n=0,1,2,3\}$ according to Eq.~\eqref{eq:ChernNumber} of the main text.

\begin{figure}
	\centering
	\includegraphics[width=\textwidth]{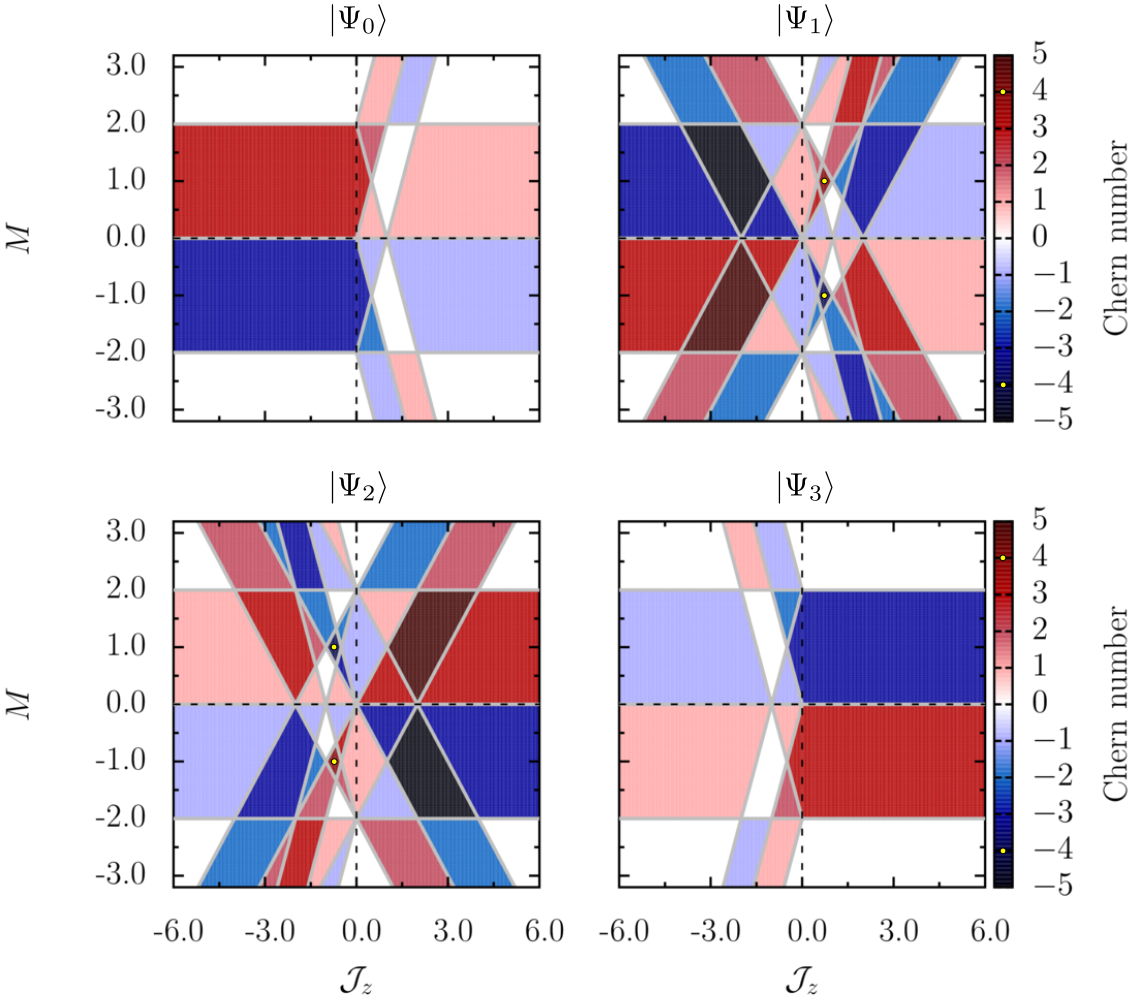}
	\caption{Topological phase diagrams for the interacting three-spin model as a function of mass parameter $M$ and effective interaction strength $\mathcal{J}_z$ ($\mathcal{J}_{x-y}=0$). Since we restrict ourselves to investigate the Hilbert subspace with $s=3/2$ (quartet states), the eigenstates of the projected Hamiltonian $\hat{H}_Q$ are given by $\ket{\Psi_n(\vec{\varphi})}$ $\{n=0,1,2,3\}$. For a better identification we have marked the phases with Chern numbers $C_n=\pm4$ by a yellow dot. Interactions drive the system into correlated topological phases with Chern numbers $C_n=\pm1,\,\pm 2,\,\pm 4,\,\pm 5$ that are forbidden in the noninteracting regime. For $C_n=\pm5$, this leads to an amplification of the topological frequency conversion that is even more pronounced than in the case of two interacting spins. 
		\label{fig:PhaseDiagram_N3_Jz}}
\end{figure}

The resulting topological phase diagrams as a function of mass parameter $M=B_0/B_c$ and effective interaction strength $\mathcal{J}_z$ ($\mathcal{J}_{x-y}=0$) are displayed in Fig.~\ref{fig:PhaseDiagram_N3_Jz}. Interactions lead to correlated topological phases with Chern numbers $C_n=\pm1,\,\pm 2,\,\pm 4,\,\pm 5$, which are completely prohibited in the noninteracting regime. Especially, the amplification of the topological frequency conversion is even more pronounced ($C_n=\pm5$) than in the case of two interacting spins. Since at HSPs the system $\hat{H}_3$ commutes with $\hat{S}_z$ (because $\mathcal{J}_{x-y}=0$), the Hamiltonian $\hat{H}_Q$ becomes diagonal in the basis of quartet states. Consequently, the topological phase transitions can be investigated in an completely analogous way as in the interacting two-spin case, which additionally illustrates that the physical interpretations of the main text are also valid for the symmetrically arranged three-spin model. 

\end{document}